\documentclass[pra,aps,amsmath,amssymb,amsfonts,nofootinbib,superscriptaddress,10pt,twocolumn]{revtex4-2}
\usepackage{bm,mathrsfs}
\usepackage{graphicx}
\usepackage{epsfig}
\usepackage{times}
\usepackage{verbatim}
\usepackage[percent]{overpic}
\usepackage{subcaption}

\usepackage[sort&compress]{natbib}
\usepackage[margin=2cm]{geometry}
\usepackage[colorlinks,breaklinks,linkcolor=blue,anchorcolor=blue,citecolor=blue,urlcolor=blue,bookmarks=false]{hyperref}
\newcommand{\ket}[1]{|#1\rangle}
\newcommand{\bra}[1]{\langle#1|}

\usepackage{lmodern}
\renewcommand{\thefootnote}{}%
\footnotetext{The paper is an English translated version of the original Chinese paper published in Acta Physica Sinica. 
	Please cite the paper as: YANG Liping, WANG Jiping, DONG Li, XIU Xiaoming, 
	JI Yanqiang, Rapid preparation of Rydberg superatom W state using superadiabatic techniques. 
	Acta Phys. Sin., 2025, 74(10): 100305. doi: 10.7498/aps.74.20241694.}
\begin{document}

\onecolumngrid   
\begin{center}
	\large{\bf \boldmath{Rapid Preparation of Rydberg Superatom W State Using Superadiabatic Techniques}}\par\vspace{5mm}
	{\rm YANG Liping$^{\#}$ \quad  WANG Jiping$^{\#}$ \quad  DONG Li \quad XIU Xiaoming \quad JI Yanqiang$^\dagger$}\par\vspace{5mm}
	\small\sl College of Physical Science and Technology,  Bohai University,  Jinzhou 121013,  China

\end{center}
\small{\narrower
	~~~The W state, as a robust multipartite entangled state, plays an important role in quantum information processing, quantum network construction and quantum computing. In this paper, we encode quantum information on the effective energy levels of Rydberg superatoms and propose a fast scheme for preparing the Rydberg superatom W state based on the superadiabatic iterative technique, this scheme can be achieved in only one step by controlling the laser pulses. In the current scheme, the superatoms are trapped in spatially separated cavities connected by optical fibers, which significantly enhances the feasibility of experimental manipulation. A remarkable feature is that it does not require precise control of experimental parameters and interaction time. Meanwhile, the form of the counterdiabatic Hamiltonian is the same as that of the effective Hamiltonian. Through numerical simulations, the fidelity of this scheme can reach 99.94$\%$. Even considering decoherence effects, including atomic spontaneous emission and photon leakage, the fidelity can still exceed 97.5$\%$, further demonstrating the strong robustness of the solution. In addition, the Rabi frequency can be characterized as a linear superposition of Gaussian functions, this representation significantly alleviates the complexity encountered in practical experiments. Futhermore, we also analyzed the impact of parameter fluctuations on the fidelity, the results show that this scheme is robust against parameter fluctuations. At last, the present scheme is extended to the cases of $N$ Rydberg superatoms, which shows the scalability of our scheme.
	
	\par}\vskip 3mm
\normalsize{\narrower{Keywords: Superadiabatic; W state; Rydberg superatoms}\par\vspace{1mm}
	
	\narrower{PACS: 03.67.Bg; 32.80.Ee; 42.50.Pq}}\par\vspace{1mm}
\noindent{\narrower{DOI: \href{http://doi.org/10.7498/aps.74.20241694}{10.7498/aps.74.20241694}}\par\vspace{1mm}
{\noindent CSTR: \href{https://cstr.cn/32037.14.aps.74.20241694}{32037.14.aps.74.20241694}}\par\vspace{5mm}
}
\twocolumngrid

\thefootnote
\section{Introduction}
With the rapid development of quantum information science and quantum computing technology, the study of multipartite entangled states has emerged as a key issue in the field of quantum information processing. Multipartite entangled states play an important role in various areas, such as quantum cryptography\cite{c01}, quantum secret sharing\cite{c02}, quantum teleportation\cite{c03, c04, c05}, quantum key distribution\cite{c06, c07, c08, c09}, quantum computation\cite{c10,c1001,c1002,c1003}and quantum secure direct communication\cite{c11, c12, c13}.

The W state is a distinctive type of multipartite entangled state\cite{c14, c03} composed of multiple qubits, Its main feature is that when any qubit is lost, the remaining qubits remain entangled with a certain probability, which makes the W state robust to local decoherence\cite{c15, c16, c17}. This characteristic endows the W state with potential applications in quantum information transmission\cite{c18, c19}, quantum key distribution\cite{c20, c09, c22} and quantum network construction\cite{c23, c24}. At present, theoretical research on the W state has become relatively mature; therefore, achieving efficient and scalable preparation of the W state is of great importance. 

In recent years, researchers have actively explored a variety of quantum systems to prepare entangled states\cite{c25,c26,c27}, among which Rydberg superatoms have emerged as a promising platform for generating quantum entangled states owing to their unique collective behavior\cite{b01, b02, b03}. The Rydberg superatom is different from a single Rydberg atom in that it is composed of an ensemble of Rydberg atoms. In the highly excited Rydberg states, the interaction strength between atoms (1–100 MHz) greatly exceeds the Rabi frequency of typical lasers (tens of kHz to several MHz)\cite{b04,beguin2013direct,xing2021realization}. Therefore, under the Rydberg blockade effect, only one atom in the Rydberg atomic ensemble within the blockade range can be excited to the Rydberg state\cite{b05, b06, b07, b08, b09,c1001,b0902, b0903, b0904}. In this case, the ensemble behaves as an effective multi-level ``atom", namely the Rydberg superatom. And the collective state of the ensemble can be use to represent the effective energy levels of the superatom\cite{b01, b10, b100,xu2021fast}. Rydberg superatoms possess many advantages\cite{b101, b02}, For example, they are easier to prepare than single atoms, and the coupling strength between the ground and excited states of a superatom is significantly stronger than in a single atom. Moreover, encoding with the energy levels of a superatom provides robustness against atom loss. Owing to these advantages, Rydberg superatoms show great potential in quantum information processing, and the preparation of quantum entangled states based on the Rydberg superatom platform has become an important research direction\cite{b11, b12}. 

So far, researchers have proposed numerous schemes to prepare entangled states with high fidelity within a short timescale\cite{b13,b14,b15}. Among the various schemes for the fast preparation of quantum entangled states, one particularly interesting method is the superadiabatic iteration technique\cite{c25,a00,xue2022implementation}, This technique is an extension of the transitionless quantum driving proposed by Berry\cite{a01}, and it was further advanced by Ibáñez et al.\cite{a02}. This method can generate a series of different Hamiltonians, allowing researchers to select those that are experimentally feasible and suitable for specific applications. Later, Song et al. applied the method to a three-level system\cite{a03}, demonstrating the potential of the superadiabatic iteration technique in multi-level settings. Subsequently Huang et al.\cite{a04}  proposed a scheme for the fast generation of GHZ states by iterating the interaction picture. while Wu et al.\cite{a05} proposed a scheme for the rapid preparation of tree-type three-dimensional entangled states based on superadiabatic techniques. Inspired by the aforementioned studies, we propose a scheme for the fast preparation of the W state based on the superadiabatic iteration technique. Different from previous methods, this scheme does not require precise control of experimental parameters or interaction times, and the counterdiabatic Hamiltonian has the same form as the effective Hamiltonian, thereby facilitating experimental implementation. In addition, this scheme requires no additional coupling, thereby ensuring high experimental feasibility while maintaining high fidelity. Furthermore, it can be extended to the rapid preparation of the W state with $N$ Rydberg superatoms. 

The structure of this paper is as follows. In section 2 introduces the physical model, presents the level structure of the Rydberg superatom, and derives the Hamiltonian under the action of the cavity field and the control field. The selection of pulse parameters for preparing the W state is then obtained by introducing the superadiabatic iteration technique. In section 3, the effectiveness of the scheme is verified by numerical simulation, and the robustness of the system under decoherence factors such as atomic spontaneous emission, cavity and fiber leakage is analyzed. The results show that the evolution time of the W state based on the superadiabatic iteration technique is shorter than that based on the adiabatic scheme. Section 4 extends the scheme to the preparation of the W state with $N$ Rydberg superatoms to explore its scalability. Finally, Section 5 summarizes the main conclusions and research significance  of this work.

\section{Physical model}
\begin{figure}[h]
	\vspace*{2mm}\vspace*{2mm}\centering
	\includegraphics[width=0.95\linewidth]{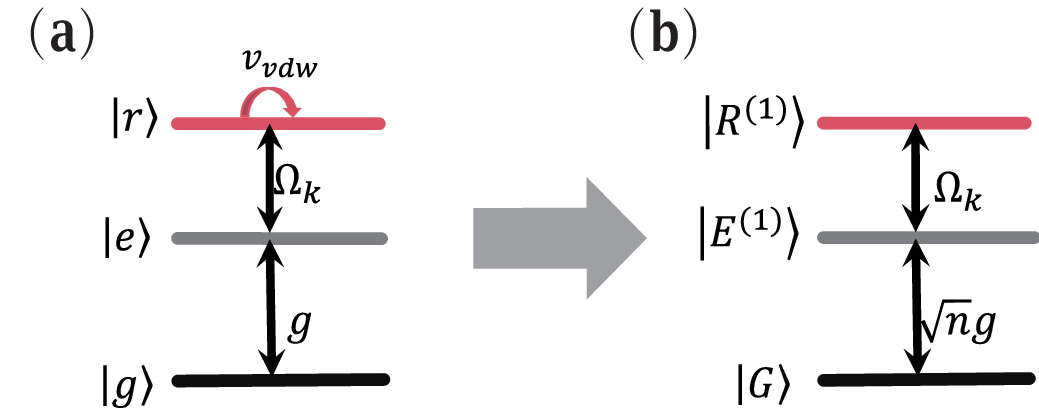}
	\caption
	{(a) Energy level structure diagram of a single Rydberg atom; (b) The equivalent energy level structure diagram of Rydberg superatom.}\label{tu1}
	\vskip 1mm
\end{figure}
\begin{figure}
	\vspace*{2mm}\centering
	\includegraphics[width=0.98\linewidth]{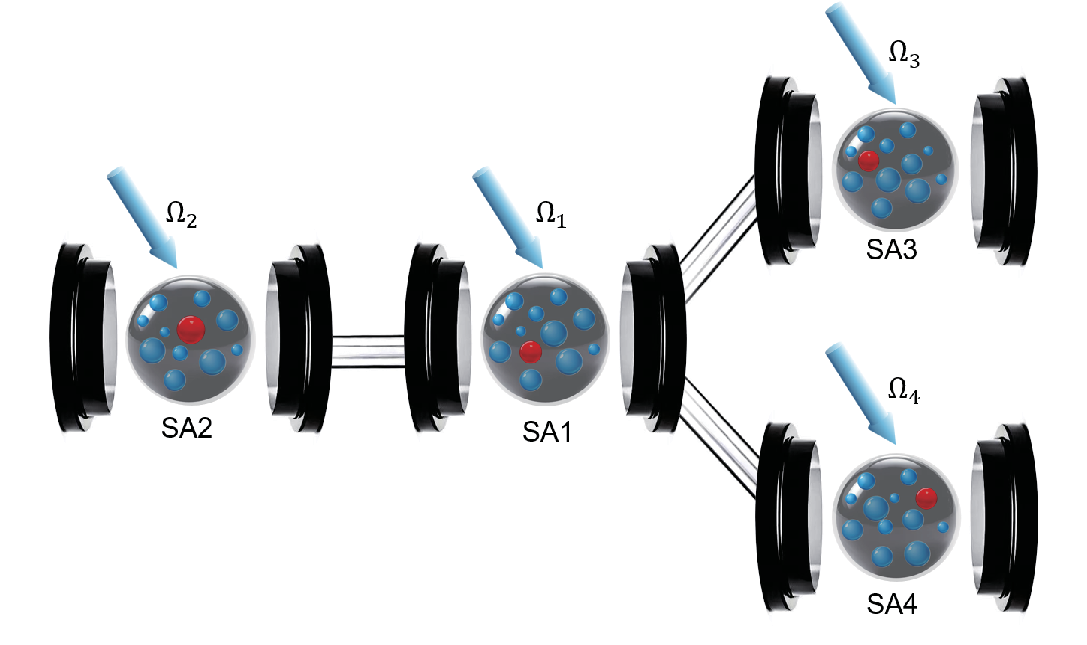}
	\caption{Schematic diagram of the structure of the Rydberg superatom-cavity system. SA denots the Rydberg superatom,  and $\Omega_k$ is the classical field Rabi frequency in the $k$-th cavity.}
	\label{tu2}
	\vskip 1mm
\end{figure}

This scheme employs 
$~^{87}\rm{Rb}$ atoms as the physical carrier of the quantum states, and its energy-level structure is shown in Fig.\ref{tu1}. An ensemble of cooled Rydberg atoms is excited under the weak cavity field limit (mean photon number does not exceed 1) and driven by an external classical field. The energy-level structure of each Rydberg atom is depicted in Fig. \ref{tu1}(a), the ground state $|g\rangle\equiv|5S_{1/2},  F=2,  m_F=2\rangle$, the excited state $|e\rangle\equiv|5P_{3/2},  F=3,  m_F=3\rangle$ and the Rydberg state $|r\rangle\equiv|111S_{1/2},  m_J=1/2\rangle$. In which the atomic transition $|g\rangle \leftrightarrow |e\rangle$ is resonantly coupled to the cavity field with a coupling constant $g$, In the $k$th cavity, the atomic transition $|e\rangle \leftrightarrow |r\rangle$ is driven by a classical field with Rabi frequency $\Omega_k(t)$, $v_{vdw}$ represents the van der Waals interaction of two atoms in a Rydberg state, which can significantly change the Rydberg energy levels of the surrounding atoms, thus forming the Rydberg blockade effect\cite{b04, t01, t02, t03, t04}. In this case, the ensemble behaves effectively as a Rydberg superatom, with the simplified energy-level structure shown in Fig. \ref{tu1} (b), $|G\rangle \leftrightarrow |E^{(1)}\rangle$ is resonantly coupled to the cavity field with a coupling constant $\lambda = \sqrt{n}g$, and the classical field resonantly drives the transition $|E^{(1)}\rangle \leftrightarrow |R^{(1)}\rangle$ of with a Rabi frequency $\Omega_k(t)$. Here,\\ $|R^{(1)}\rangle=\sum^n_i|g_1g_2...r_i...g_n\rangle/\sqrt{n}$,\\ $|E^{(1)}\rangle=\sum_i^n|g_1g_2...e_i...g_n\rangle/\sqrt{n}$\\$|G\rangle=|g_1g_2...g_n\rangle$. \\
The superatoms are placed in four independent vacuum cavities, as shown in fig. \ref{tu2}. 
Under the rotating wave approximation, the Hamiltonian in the interaction picture can be expressed as($\hbar=1$):
\begin{align}
	H_{tot}&=H_{ac}+H_{al}+H_{cf},  \cr H_{ac}&=\sum^4_{k=1}\lambda_{k}a_{k}|E^{(1)}\rangle_k\langle{G}|+{\rm H.c.}, \cr   
	H_{al}&=\sum^4_{k=1}\Omega_{k}(t)|E^{(1)}\rangle_{k}\langle{R^{(1)}}|+{\rm H.c.}, \cr
	H_{cf}&=\sum^3_{k=1}v(a_{1}+a_{k+1})b_{k}^\dag+{\rm H.c.}, 
	\label{Eq.1}
\end{align}
Where, $H_{ac}$ denotes the interaction between the cavity and the superatom, $H_{al}$ denotes the interaction between the classical field and the super-atom, and $H_{cf}$ represents the interaction of the cavity and the fiber. $a_k$ is the annihilation operator of the photon in the  $k$th cavity, $v$ represents the coupling strength between the fiber and the cavity, and $b_{k}^\dag$is the creation operator of the $k$th fiber. The initial state of the system is assumed to be $|\psi_1\rangle=|R^{(1)}GGG\rangle|0\rangle_{all}$, that is, the first superatom is in the state $|R^{(1)}\rangle$, the other superatoms are in the ground state $|G\rangle$, and all cavities and fibers are in the vacuum state. Under this assumption, the system will evolve in the space as follows
\begin{equation}
	\begin{array}{ll}  
		|\psi_1\rangle=|R^{(1)}GGG\rangle|0\rangle_{all},  & |\psi_2\rangle=|E^{(1)}GGG\rangle|0\rangle_{all},  \\ |\psi_3\rangle=|GGGG\rangle|1\rangle_{c1},  &
		|\psi_4\rangle=|GGGG\rangle|1\rangle_{f_1},   \\  |\psi_5\rangle=|GGGG\rangle|1\rangle_{c2},   &  |\psi_6\rangle=|GE^{(1)}GG\rangle|0\rangle_{all}, \\
		|\psi_7\rangle=|GR^{(1)}GG\rangle|0\rangle_{all},   &  |\psi_8\rangle=|GGGG\rangle|1\rangle_{f_2},   \\  |\psi_9\rangle=|GGGG\rangle|1\rangle_{c_3},   &
		|\psi_{10}\rangle=|GGE^{(1)}G\rangle|0\rangle_{all},   \\  |\psi_{11}\rangle=|GGR^{(1)}G\rangle|0\rangle_{all},   &  |\psi_{12}\rangle=|GGGG\rangle|1\rangle_{f_3},  \\
		|\psi_{13}\rangle=|GGGG\rangle|1\rangle_{c_4},   &  |\psi_{14}\rangle=|GGGE^{(1)}\rangle|0\rangle_{all},   \\ |\psi_{15}\rangle=|GGGR^{(1)}\rangle|0\rangle_{all}.
	\end{array}
	\label{Eq.2}
\end{equation}
In this space, the Hamiltonian can be rewritten as
\begin{align}
	H_{tot} =& H_{ac} + H_{al} + H_{cf}, \cr
	H_{ac} =& \lambda \big( |\psi_3\rangle \langle \psi_2|
	+ |\psi_6\rangle \langle \psi_5|
	+ |\psi_{10}\rangle \langle \psi_9|
	\cr &+ |\psi_{14}\rangle \langle \psi_{13}| \big) + {\rm H.c.}, \cr
	H_{al} =& \Omega_1(t)|\psi_2\rangle \langle \psi_1|
	+ \Omega_2(t)|\psi_7\rangle \langle \psi_6|
	\cr &+ \Omega_3(t)|\psi_{11}\rangle \langle \psi_{10}|
	+ \Omega_4(t)|\psi_{15}\rangle \langle \psi_{14}| + {\rm H.c.}, \cr
	H_{cf} =& v \big( |\psi_4\rangle \langle \psi_3|
	+ |\psi_5\rangle \langle \psi_4|
	+ |\psi_8\rangle \langle \psi_3|
	\cr &+ |\psi_9\rangle \langle \psi_8|
	+ |\psi_{12}\rangle \langle \psi_3|
	+ |\psi_{13}\rangle \langle \psi_{12}| \big) + {\rm H.c.}.
	\label{Eq.3}
\end{align}
For simplicity, it is assumed that 
$v=\lambda$ in the calculation. According to quantum Zeno dynamics, the Hilbert space can be decomposed into several subspaces. Since the initial state is
$|\psi_1\rangle=|R^{(1)}GGG\rangle|0\rangle_{all}$, the whole evolution process will be restricted to the dark-state subspace spanned by \{$|\psi_{1}\rangle$, $|\psi_{7}\rangle$, $|\psi_{11}\rangle$, $|\psi_{15}\rangle$, $|\phi\rangle$\} , where $|\phi\rangle=1/\sqrt{15}(3|\psi_2\rangle-|\psi_{4}\rangle+|\psi_{6}\rangle-|\psi_{8}\rangle+|\psi_{10}\rangle-|\psi_{12}\rangle+|\psi_{14}\rangle)$. Thus, the effective Hamiltonian can be written as
\begin{align}
	H_{eff}=\xi_0P_0+P_0H_{al}P_0, 
	\label{Eq.4}
\end{align}
where $P_0$ denotes the projection operator\cite{t0}, The explicit expression is $ P_0=\sum_m|m\rangle\langle m| ~(m\in\{|\psi_1\rangle, |\psi_7\rangle, |\psi_{11}\rangle, |\psi_{15}\rangle, |\phi\rangle\})$. Since the system evolves in dark space, Since the system evolves in dark space, i.e. $\xi_0=0$. We let $\Omega'_2(t)=1/\sqrt{5}\Omega_2(t)=1/\sqrt{5}\Omega_3(t)=1/\sqrt{5}\Omega_4(t), ~\Omega'_1(t)=3\Omega_1(t)/\sqrt{15}$, $|\Psi\rangle=1/\sqrt{3}(|\psi_7\rangle+|\psi_{11}\rangle+|\psi_{15}\rangle), $ then the effective Hamiltonian becomes
\begin{align}
	H_{eff}=\Omega'_1(t)|\phi\rangle\langle\psi_1|+\Omega'_2(t)|\Psi\rangle\langle\phi|+{\rm H.c.}.
	\label{Eq.5}
\end{align}
For convenience, let $\left|\psi_1\right\rangle=\left(1, 0, 0\right)^T$,  $\left|\phi\right\rangle=\left(0, 1, 0\right)^T$ and
$\left|\Psi\right\rangle=\left(0, 0, 1\right)^T$,  in which case the effective Hamiltonian can be rewritten as
\begin{align}
	H_{eff}=  \Omega'(t) \begin{pmatrix}
		0 & \sin \theta_1(t) & 0 \\
		\sin \theta_1(t) & 0 & \cos \theta_1(t) \\
		0 & \cos \theta_1(t) & 0
	\end{pmatrix}, 
	\label{Eq.6}
\end{align}
where, 
\begin{align}
	\theta_1(t)=\arctan \frac{\Omega'_1(t)}{\Omega'_2(t)}, ~~~~\Omega'(t)=\sqrt{\Omega'_1(t)^2+\Omega'_2(t)^2}. 
	\label{Eq.7}
\end{align}
The eigenvalues of this effective Hamiltonian are $\zeta_0 = 0$ and  $\zeta_\pm = \pm \Omega'(t)$, and the corresponding eigenstates are
\begin{align}
	&\left|\varphi_0(t)\right\rangle=\left(\begin{array}{c}
		\cos \theta_1(t) \\
		0 \\
		-\sin \theta_1(t)
	\end{array}\right), \cr
	&\left|\varphi_\pm(t)\right\rangle=\frac{1}{\sqrt{2}}\left(\begin{array}{c}
		\sin \theta_1(t) \\
		\pm1 \\
		\cos \theta_1(t)
	\end{array}\right).
	\label{Eq.8}
\end{align}
According to the adiabatic evolution theory\cite{wu2016adiabatic,shore2017picturing}, if the initial state of a system is in an eigenstate, it can continue to evolve adiabatically along the eigenstate. Specifically, if the state of the system at the initial time is $|\psi_1\rangle$, and is in 
the eigenstate $\zeta_0(t)$, then the system will evolve adiabatically along the eigenstate $\zeta_0(t)$. The target state of the scheme is the W state i.e. $1/2 (|\psi_1\rangle + |\psi_7\rangle + |\psi_{11}\rangle + |\psi_{15}\rangle)$, which can be further expressed as $1/2 |\psi_1\rangle + \sqrt{3}/2 |\Psi\rangle$. According to this, it is easy to obtain the boundary conditions about $\theta_1$, that is, $\theta_1(0)=0$, $\theta_1(T)=-\pi/3$ (where $T$ is the evolution time of the system). For a better analysis of the Hamiltonian, a unitary transformation is applied to it
\begin{align}
	A_1 = \begin{pmatrix}
		\cos \theta_1(t) & \frac{\sin \theta_1(t)}{\sqrt{2}} & \frac{\sin \theta_1(t)}{\sqrt{2}} \\
		0 & \frac{1}{\sqrt{2}} & -\frac{1}{\sqrt{2}} \\
		-\sin \theta_1(t) & \frac{\cos \theta_1(t)}{\sqrt{2}} & \frac{\cos \theta_1(t)}{\sqrt{2}}
	\end{pmatrix}, 
\end{align}
through $H_1=A_1^{\dag}H_{eff}A-i A_1^{\dag} \partial_t A_1$, we can obtain
\begin{align}\label{Eq.10}
	H_1=&\begin{pmatrix}
		0 & 0 & 0  \\
		0 & \Omega'(t) & 0  \\
		0 & 0 & -\Omega'(t) 
	\end{pmatrix}\cr&-i\begin{pmatrix}
		0 & \frac{\dot{\theta}_1(t)}{\sqrt{2}} & \frac{\dot{\theta}_1(t)}{\sqrt{2}} \\
		-\frac{\dot{\theta}_1(t)}{\sqrt{2}} & 0 & 0 \\
		-\frac{\dot{\theta}_1(t)}{\sqrt{2}} & 0 & 0
	\end{pmatrix}, 
\end{align}
According to (\ref{Eq.10}), we can found the adiabatic approximation condition for adiabatic evolution is $|\Omega'(t)|\gg {|\dot{\theta}_1}/{\sqrt{2}}|$, this is, coupling strength $|{\dot{\theta}_1}/{\sqrt{2}}|$ between different eigenstates must be negligible. Some researchers have proposed adding a counterdiabatic term to eliminate or suppress this adiabatic coupling\cite{a05}. It is easy to see that adding the term $H^{(1)}_{CD}=i A_1^{\dag} \partial_t A_1$ to the Hamiltonian can cancel exactly the term that causes the adiabatic coupling $-i A_1^{\dag} \partial_t A_1$. Therefore, the counterdiabatic term added to the original Hamiltonian can be written as
\begin{align}\label{Eq.11}
	H_{CD}^{(1)}=\begin{pmatrix}
		0 & 0 & i\dot{\theta}_1(t)  \\
		0 & 0 & 0  \\
		-i\dot{\theta}_1(t)  & 0 & 0
	\end{pmatrix}, 
\end{align}
However, Eq.(\ref{Eq.11}) indicates that the added counterdiabatic Hamiltonian requires a direct coupling between $\ket{\psi_1}$ and $\ket{\Psi}$, which is generally difficult to implement in practice. In order to find the appropriate Hamiltonian, the analysis is carried out in the superadiabatic picture (i.e., the second picture iteration). The eigenvalues of the Hamiltonian $H_{1}$ (\ref{Eq.10}) are 
$ 0,~\pm \Omega'^\prime(t)$,  where $\Omega'^\prime(t)=\sqrt{\Omega'^2(t)+\dot{\theta}_1^2(t)}$. The corresponding eigenstates are
\begin{align}
	&\left|\xi_0(t)\right\rangle=\left(\begin{array}{c}
		-i\sin \theta_2(t) \\
		-\frac{\cos \theta_2(t)}{\sqrt{2}} \\
		\frac{\cos \theta_2(t)}{\sqrt{2}}
	\end{array}\right),\cr 
	&\left|\xi_\pm(t)\right\rangle=\frac{1}{\sqrt{2}}\left(\begin{array}{c}
		\mp i\frac{\cos \theta_2(t)}{\sqrt{2}} \\
		\frac{1}{2}(1\pm \sin \theta_2(t)) \\
		\frac{1}{2}(1\mp \sin \theta_2(t))
	\end{array}\right), 
\end{align}
in the formula  $\theta_2(t)=\arctan |\dot{\theta}_1(t)|/\Omega'(t)$. Next, we perform the second unitary transformation
\begin{align}
	A_2=\begin{pmatrix}
		-i \sin \theta_2(t) & -\frac{i \cos \theta_2(t)}{\sqrt{2}} & \frac{i \cos \theta_2(t)}{\sqrt{2}} \\
		-\frac{\cos \theta_2(t)}{\sqrt{2}} & \frac{1}{2}(1 + \sin \theta_2(t)) & \frac{1}{2}(1 - \sin \theta_2(t)) \\
		\frac{\cos \theta_2(t)}{\sqrt{2}} & \frac{1}{2}(1 - \sin \theta_2(t)) & \frac{1}{2}(1 + \sin \theta_2(t))
	\end{pmatrix}
\end{align}
through $H_2=A_2^{\dag}H_1 A-i A_2^{\dag} \partial_t A_2$,  it can be easily obtained
\begin{align}
	H_2=&\begin{pmatrix}
		0 & 0 & 0  \\
		0 & \Omega'^\prime(t) & 0  \\
		0 & 0 & -\Omega'^\prime(t) 
	\end{pmatrix}
	\cr&-i\begin{pmatrix}
		0 & -\frac{\dot{\theta}_2(t)}{\sqrt{2}} & \frac{\dot{\theta}_2(t)}{\sqrt{2}} \\
		\frac{\dot{\theta}_2(t)}{\sqrt{2}} & 0 & 0 \\
		-\frac{\dot{\theta}_2(t)}{\sqrt{2}} & 0 & 0
	\end{pmatrix}, 
	\label{Eq.14}
\end{align}
the counterdiabatic term in the superadiabatic picture can be obtained as $H^{(2)}_{CD}=i A_2^{\dag} \partial_t A_2$, which corresponds to the second term in (\ref{Eq.14}). Its explicit form in the original picture can be readily obtained as $H^{(2)}_{CD}=i A_1 \partial_t A_2 A_2^{\dag} A_1^{\dag}$
\begin{align}\label{Eq.15}
	H_{CD}^{(2)}=\dot{\theta}_2(t)\begin{pmatrix}
		0 & -\cos \theta_1(t) & 0  \\
		-\cos \theta_1(t) & 0 & \sin \theta_1(t)  \\
		0  & \sin \theta_1(t) & 0
	\end{pmatrix}.
\end{align}
By adding the counterdiabatic term to the effective Hamiltonian 
$H_{eff}$, the modified Hamiltonian is given by
\begin{widetext}
	\begin{align}
		\widetilde{H}=& H_{eff}+H_{CD}^{(2)}\cr
		=& \begin{pmatrix}
			0 &  \Omega'(t)\sin \theta_1(t)-\dot{\theta}_2(t) \cos \theta_1(t) & 0  \\
			\Omega'(t)\sin \theta_1(t)- \dot{\theta}_2(t) \cos \theta_1(t) & 0 & \Omega'(t)\cos \theta_1(t)+\dot{\theta}_2(t) \sin \theta_1(t)  \\
			0  & \Omega'(t)\cos \theta_1(t)+\dot{\theta}_2(t) \sin \theta_1(t) & 0
		\end{pmatrix}, 
	\end{align}
\end{widetext}
It can be seen that the added counterdiabatic term $H_{CD}^{(2)}$ is a correction to the original pulse, which can be readily implemented in experiments. 

\section{Numerical simulation}
\begin{figure}
	\centering
	\makebox[\linewidth]{
	\includegraphics[width=1.6\linewidth]{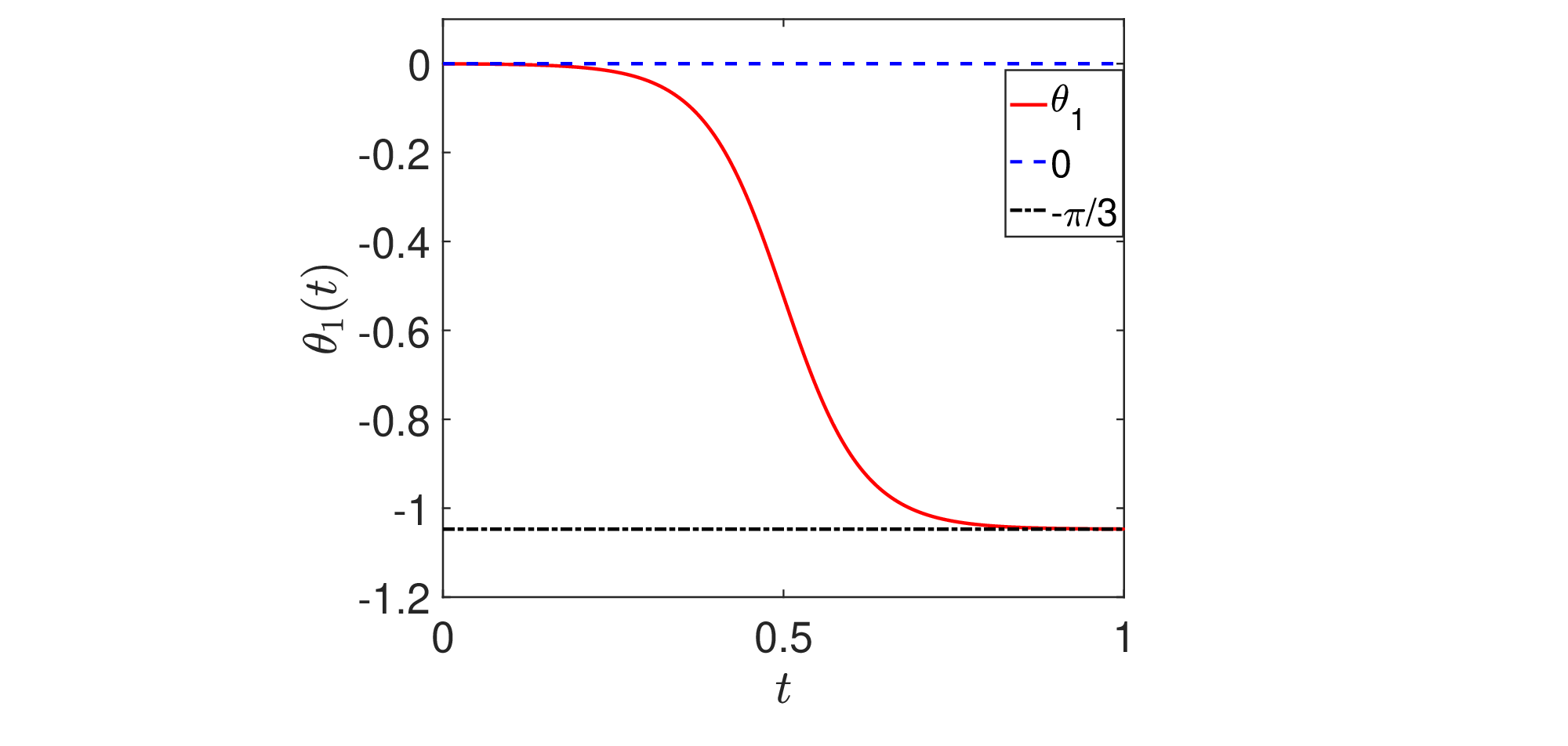}
	}
	\caption{The $\theta_1(t)$ as a function of the time.
		The parameters $t_0 = 0.14T $ and $t_c = 0.19T $.}
	\label{tu3}
\end{figure}

\begin{figure}
	\centering
	\makebox[\linewidth]{
	\includegraphics[width=1.4\linewidth]{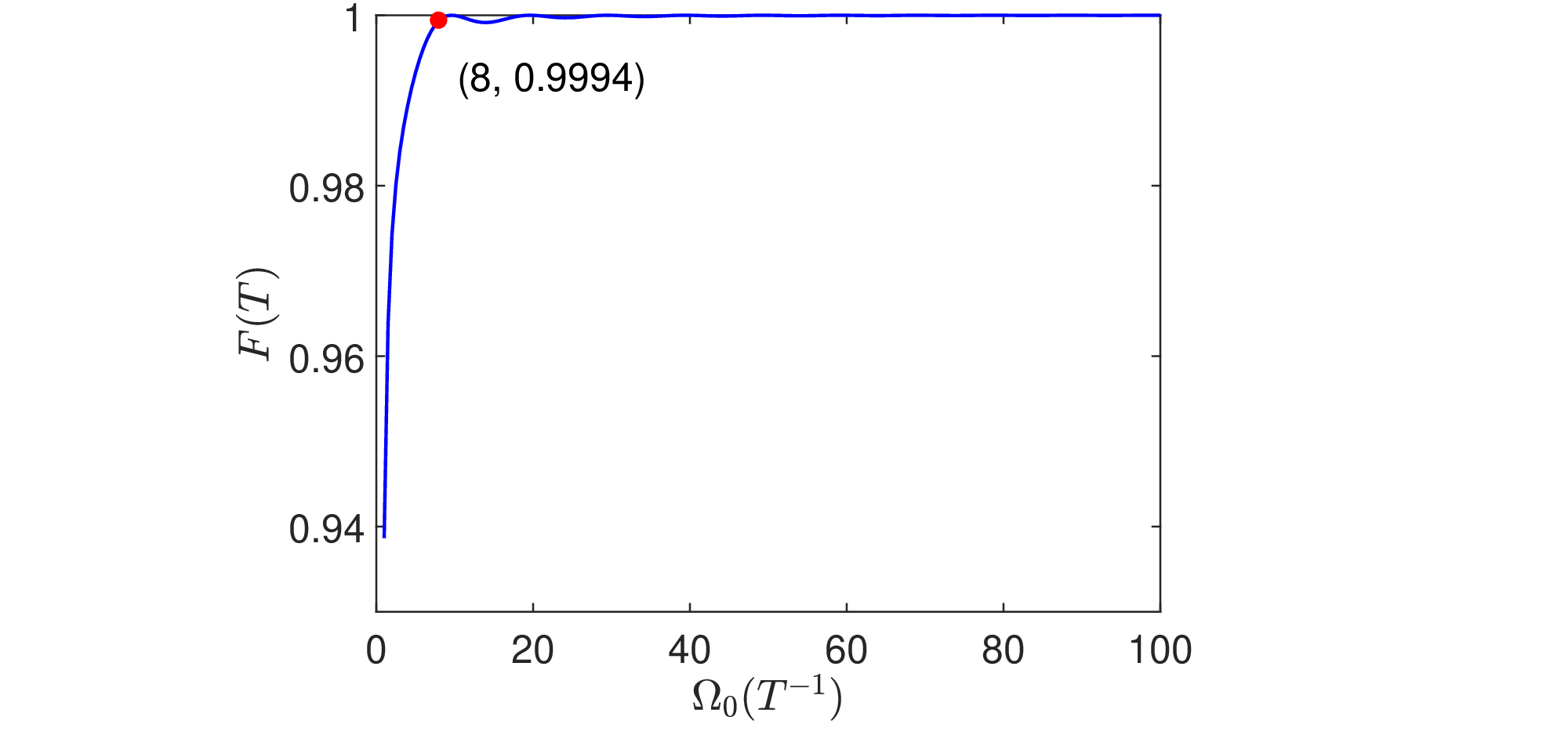}
	}
	\caption{The influence of $\Omega_0(T^{-1})$ on fidelity $F(T)$. When $\Omega_0=8T^{-1}$,  the fidelity $F(T)=0.9994$.}
	\label{tu5}
	\vskip 1mm
\end{figure}

According to the boundary conditions of the stimulated Raman adiabatic passage,  $\Omega^{\prime}_{1}$ and $ \Omega^{\prime}_{2}$ can be chosen in the following form
\begin{align}
	\Omega^{\prime}_{1}(t)=&\sin{\vartheta}\Omega_{0}\exp\left[\frac{-(t-t_{0}-T/2)^2}{t^2_{c}}\right], \cr
	\Omega^{\prime}_{2}(t)=&\Omega_{0}\exp\left[\frac{-(t+t_{0}-T/2)^2}{t^2_{c}}\right]\cr &+\cos{\vartheta}\Omega_{0}\exp\left[\frac{-(t-t_{0}-T/2)^2}{t^2_{c}}\right].
	\label{Eq.17}
\end{align}
where, $ \vartheta=-\pi/3$,  $\Omega_0$ is the pulse amplitude, here $t_0=0.14T$, $t_c=0.19 T$, $T$ is the evolution time. For these two pulses, one can determine the $\theta_1(t)$ and $\theta_2(t)$. In Fig.~\ref{tu3}, we plot the relation between 
$\theta_1(t)$ and the evolution time. According to (\ref{Eq.7}) and (\ref{Eq.17}),  $\theta_1(t)$ is independent of $\Omega_0$. 
And by combining with Fig.~\ref{tu3}, it can be concluded that, regardless of the value of 
$\Omega_0$, the conditions 
$\theta_1(0) = 0$ and 
$\theta_1(T) = -\pi/3$ are always satisfied.
In Fig.~\ref{tu3}, we plot the relation between 
$|\theta_2(t)|$ and the evolution time, we can found when $\Omega_0$  is large enough, the condition  $\theta_2(0) = \theta_2(T)$ can also be satisfied. However, since the limit condition $\Omega_k(t) \ll \{\lambda,v\}_{min}$ is required in this scheme, so  the  $\Omega_0$ cannot be too large.
\begin{figure}
	\centering
	\makebox[\linewidth]{
	\includegraphics[width=1.7\linewidth]{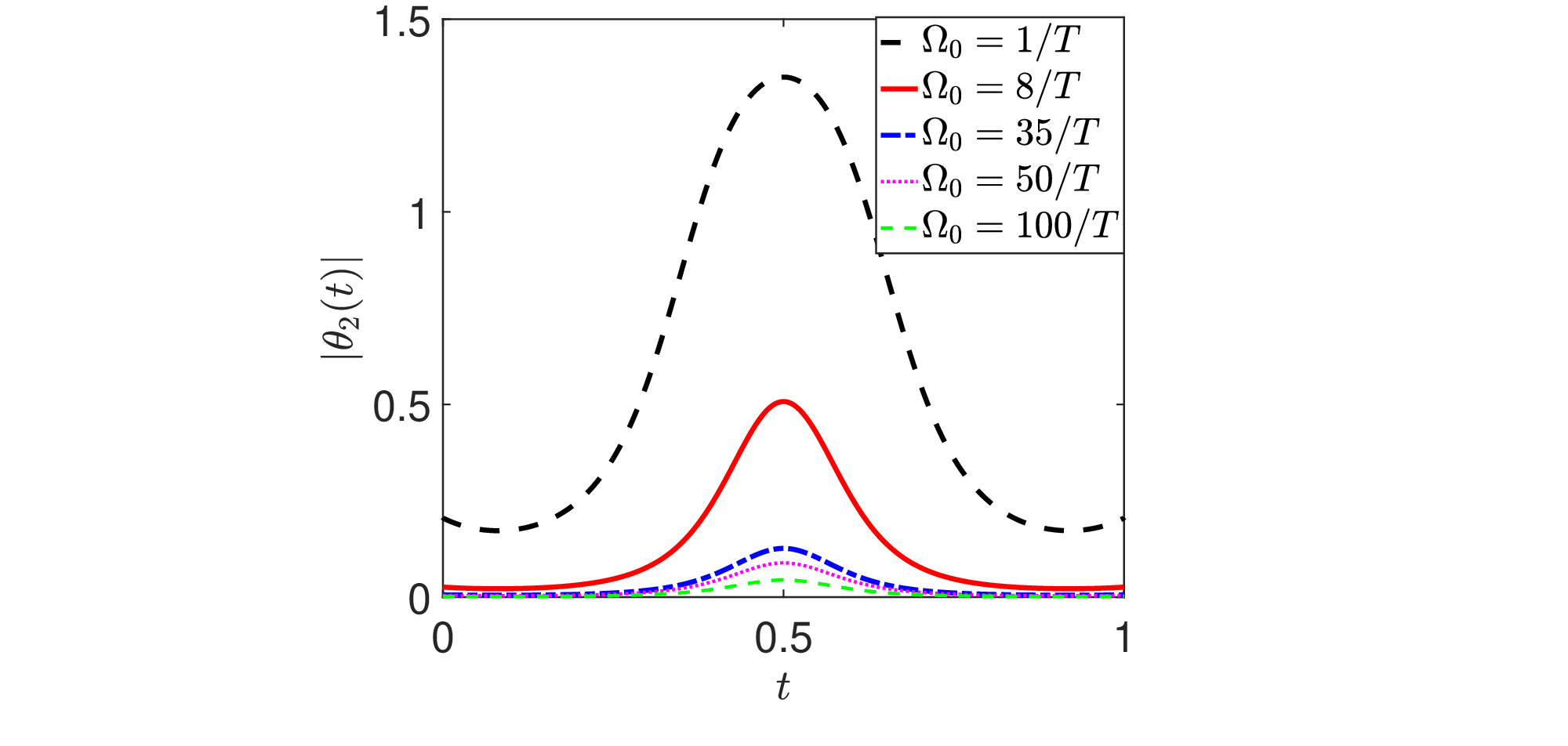}
	}
	\caption{The $\theta_2(t)$ as a function of the time. The parameters $t_0 = 0.14T$and $t_c = 0.19T$.}
	\label{tu4}
\end{figure}
In order to select the appropriate $\Omega_0$ , we plot the final fidelity Fig. \ref{tu5} plots  $F(T) = |\langle{\psi_W}|\psi(T)\rangle|^2$ versus $\Omega_0$ , where $\psi(T)$  is the final state of the whole system. According to Fig.~\ref{tu5}, one can choose 
$\Omega_0=8/T$ and 
$\lambda = 80/T$.

\begin{figure}
	\centering
	\makebox[\linewidth]{
	\includegraphics[width=1.6\linewidth]{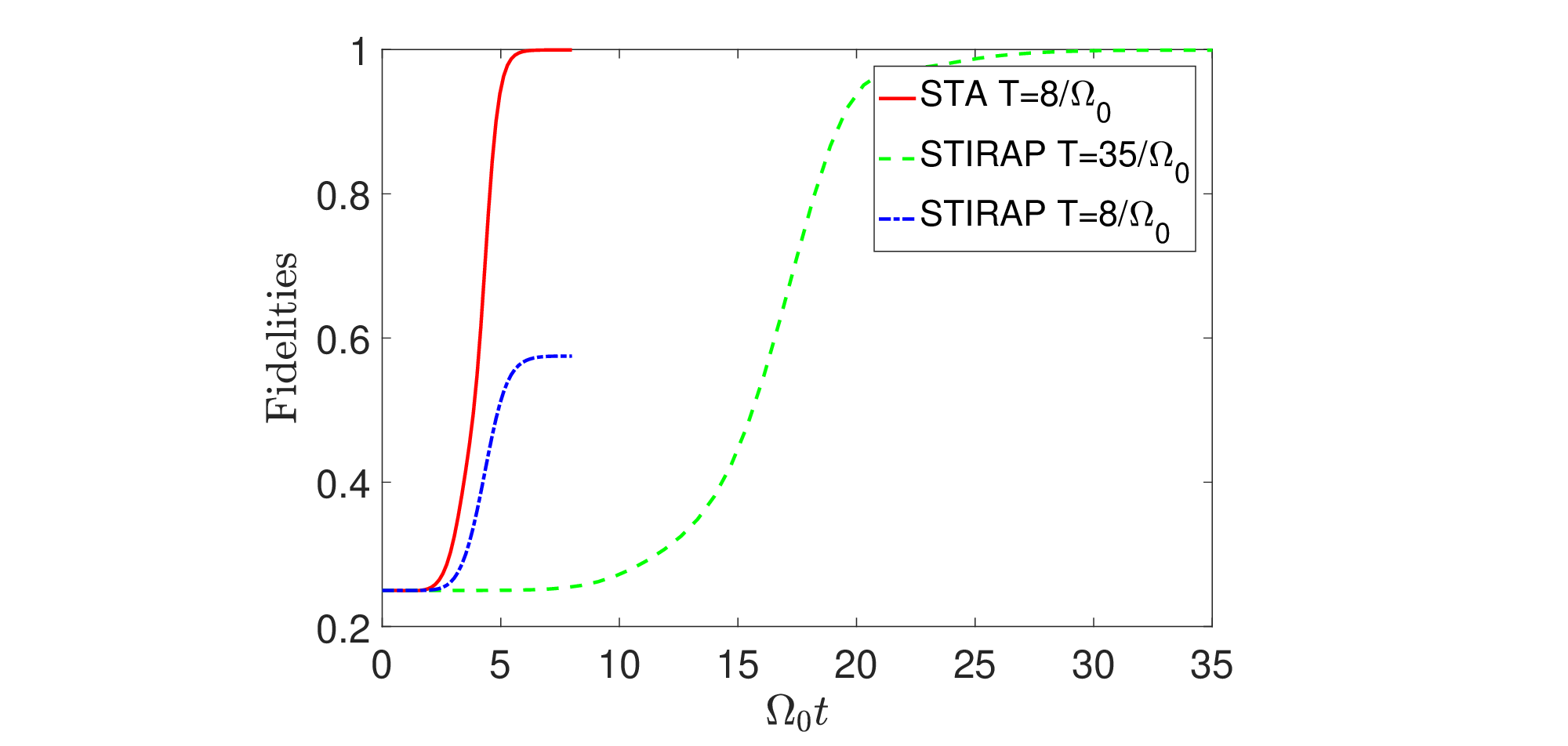}}
	\caption{Under the three different conditions: superadiabatic iteration $T=8/\Omega_0$, adiabatic evolution $T=8/\Omega_0$ and adiabatic evolution $T=35/\Omega_0$, the fidelity of W state as a function of the time.}
	\label{tu7}
	\vskip 1mm
\end{figure}

Due to the complexity of the $\Omega'_1(t)$ and $\Omega'_2(t)$, Gaussian fitting is applied to further enhance the experimental feasibility. The two fitted pulses can be expressed as
\begin{align}
	&\widetilde{\Omega}_1(t)=\sum_{i=1}^2 c_{1i}\exp^{-(t-m_{1i})^2/n^2_{1i}}, \cr &
	\widetilde{\Omega}_2(t)=\sum_{i=1}^2 -c_{2i}\exp^{-(t-m_{2i})^2/n^2_{2i}}, 
\end{align}
The corresponding parameters of the two pulses are
\begin{center}
	\begin{tabular}{llll} 
		$c_{11}=5.912/T, $ & $m_{11}=0.6838T, $ & $n_{11}=0.1561T, $ \cr $c_{12}=4.784 /T, $ & $m_{12}= 0.4265T, $ & $n_{12}=0.09342T, $ \cr  $c_{21}=7.590/T, $ & $m_{21}=0.5857T, $ & $n_{21}=0.1888T, $ \cr  $c_{22}=7.111/T, $ & $m_{22}= 0.3132T, $ & $n_{22}=0.1538T.$ &
	\end{tabular}
\end{center}
\begin{figure}
	\centering
	\makebox[\linewidth]{
	\includegraphics[width=1.3\linewidth]{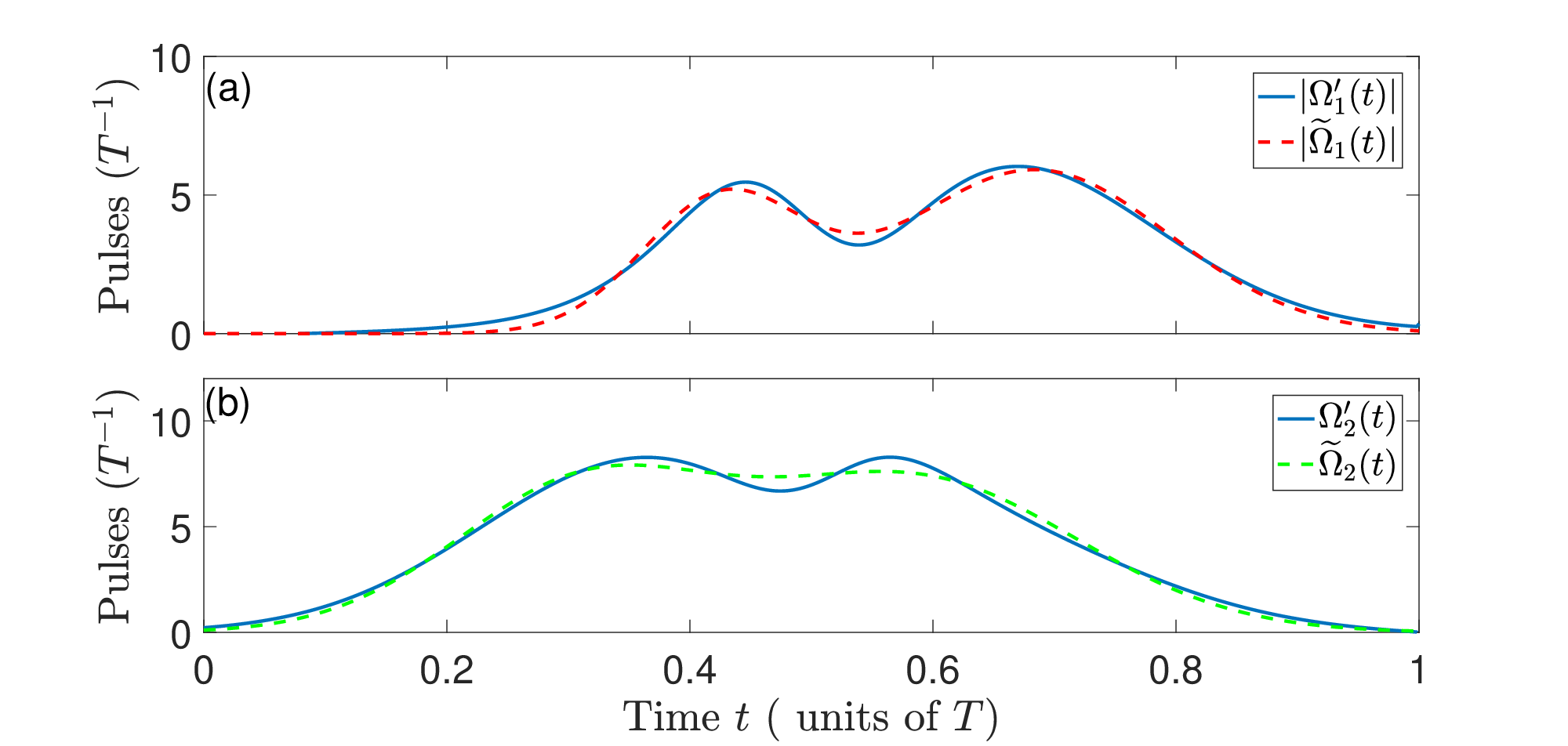}}
	\caption{(a) Comparing the pulse $\Omega '_1 (t) $ and the fitting of gaussian pulse $\widetilde {\Omega} _1 (t) $. (b) Comparing the pulse $\Omega '_2 (t) $ and the fitting of gaussian pulse $\widetilde {\Omega} _2 (t) $.}
	\label{tu6}
	\vskip 1mm
\end{figure}
As shown in Fig. \ref{tu6},  $\widetilde{\Omega}_1(t)$ and $\widetilde{\Omega}_2(t)$ of the fitted pulse are highly coincident with the  $\Omega'_1(t)$ and $\Omega'_2(t)$of the original pulse, so the fitted pulse can be used to replace the original pulse for test operation.  
To illustrate the features of our scheme, Fig. \ref{tu7} shows the time evolution of the fidelity of the W state under different conditions. By comparing the three cases, it can be concluded that the superadiabatic iteration method requires significantly less time to evolve to the W state than the adiabatic method.

\begin{figure}
	\centering
	\makebox[\linewidth]{
	\includegraphics[width=1.2\linewidth]{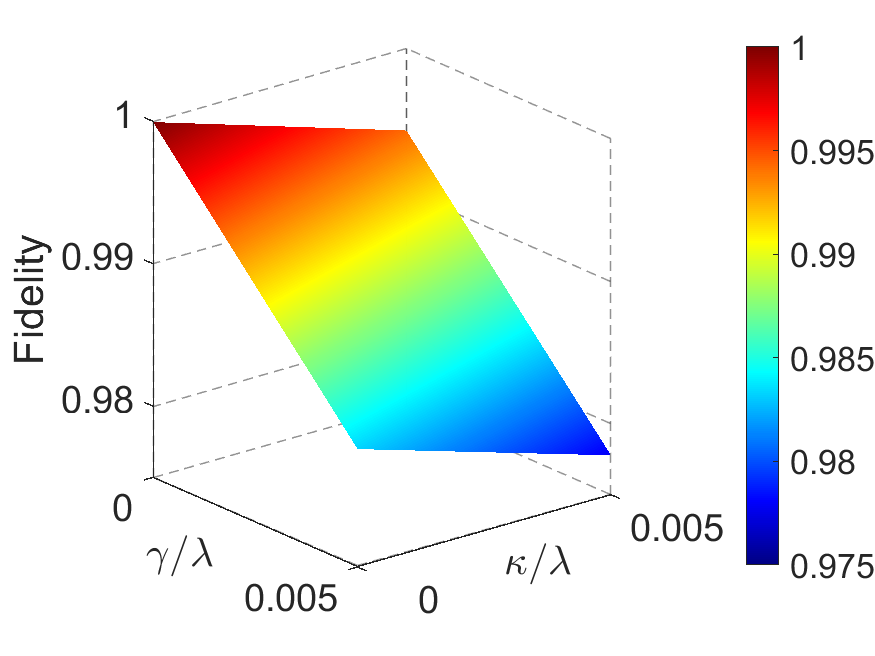}}
	\caption{The relationship between the fidelity of the W state and $\kappa/\lambda$, $\gamma/\lambda$ by the Hamiltonian $H_{tot}$. $T = 8 / \Omega_0, ~\Omega_0 = 0.1\lambda$.}
	\label{tu8}
	\vskip 1mm
\end{figure}

\begin{figure}
	\centering
	\makebox[\linewidth]{
		\includegraphics[width=1.0\linewidth]{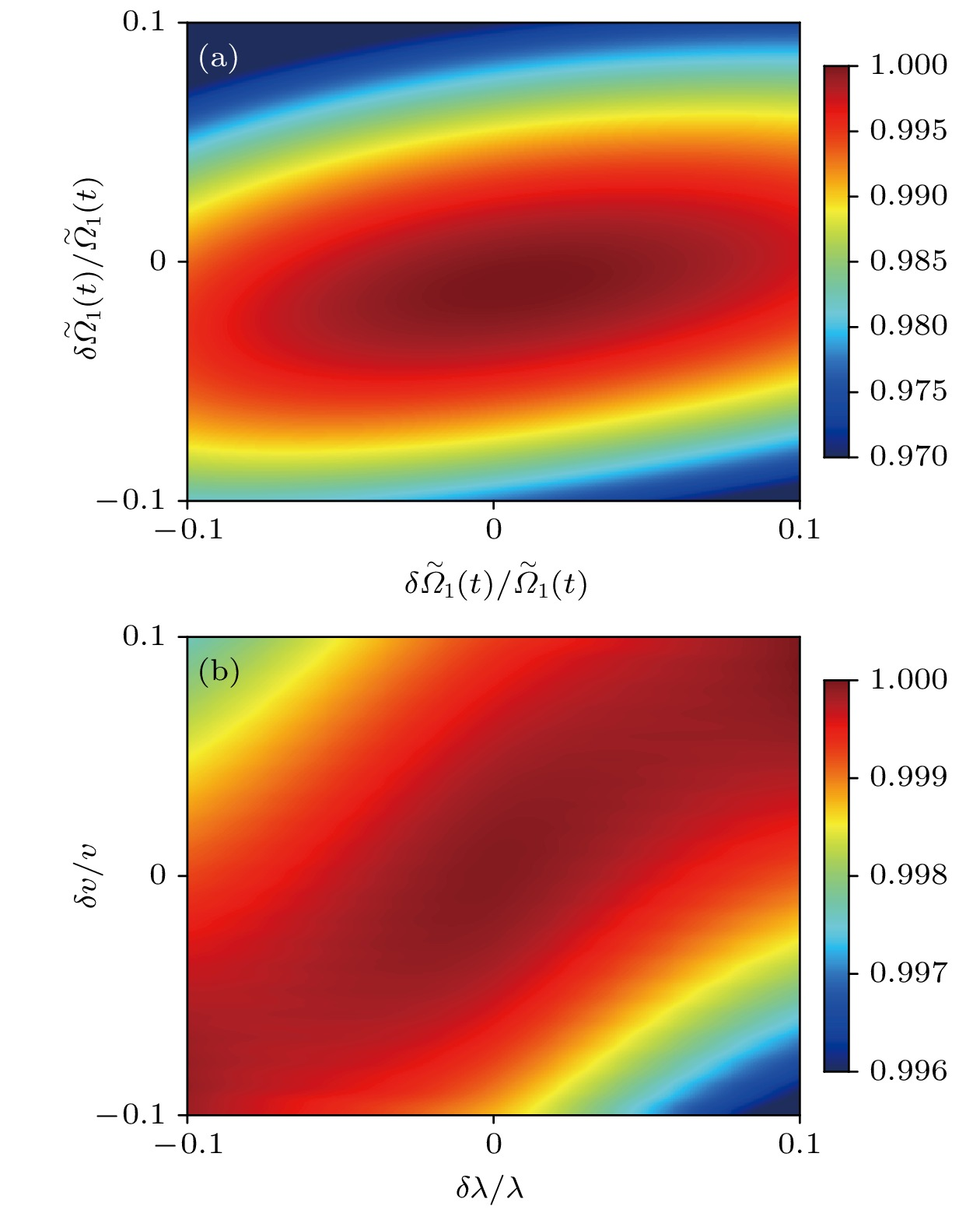}}
	\caption{(a)The fidelity versus $\delta \widetilde{\Omega}_1$ and $\delta \widetilde{\Omega}_2$. (b)The fidelity versus $\delta \lambda$ and $\delta v$.}
	\label{tu9}
\end{figure}

In addition, in order to further verify the practical effect of the scheme, the effects of atomic spontaneous emission, photon leakage in the cavity and fiber on the scheme are also considered, and the master equation of the system can be written as
\begin{align}
	\dot{\rho}(t)=&i[\rho(t), H_{tot}]\cr &+\sum_{k=1}^4 \left[L_k^{R} \rho L_k^{R}~^\dag-\frac{1}{2}(L_k^{R}~^\dag L_k^{R}~\rho+\rho L_k^{R}~^\dag L_k^{R}~)\right] \cr&+\sum_{k=1}^4 \left[L_k^{E} \rho L_k^{E}~^\dag-\frac{1}{2}(L_k^{E}~^\dag L_k^{E}~\rho+\rho L_k^{E}~^\dag L_k^{E}~)\right]\cr
	&+\sum_{k=1}^4 \left[L_k^{c} \rho L_k^{c}~^\dag-\frac{1}{2}(L_k^{c}~^\dag L_k^{c}~\rho+\rho L_k^{c}~^\dag L_k^{c}~)\right]\cr
	&+\sum_{k=1}^3 \left[L_k^{f} \rho L_k^{f}~^\dag-\frac{1}{2}(L_k^{f}~^\dag L_k^{f}~\rho+\rho L_k^{f}~^\dag L_k^{f}~)\right],\cr
	~
\end{align}
where, $L_k$describes the decoherence effect, $L_k^{R}=\sqrt{\gamma_k^{R}}\ket{E^{(1)}}_k\bra{R^{(1)}}$ is $k$-th superatom  $\ket{R^{(1)}}\rightarrow \ket{E^{(1)}}$, 
$L_k^{E}=\sqrt{\gamma_k^{E}}\ket{G}_k\bra{E^{(1)}}$ represents the spontaneous emission of the 
$k$-th superatom from $\ket{E^{(1)}}$ to $ \ket{G}$.
$L_k^{c}=\sqrt{\kappa_k^{c}}a$ is the attenuation of photons in the $k$-th cavity, and $L_k^{f}=\sqrt{\kappa_k^{f}}b$ is the attenuation of photons in the $k$-th fiber. For ease of calculation, let $\gamma_k^{R}=0.01\gamma,~\gamma_k^{E}=\gamma$, and $\kappa_k^{f}=\kappa_k^{c}=\kappa$. 
It can be seen from the Fig. \ref{tu8} that the superadiabatic scheme is more sensitive to the atomic spontaneous emission than the photon leakage in the cavity-fiber system. But in general, the scheme is robust to the decoherence effects caused by atomic spontaneous emission and photon leakage
Even in the case of $\kappa/\lambda=\gamma/\lambda=0.005$, the final fidelity $F(T)$ is still higher than $97.5\%$. 
In practical operations, operational deviations are inevitable. Therefore, we further investigate the influence of variations in the fitted pulses $\widetilde{\Omega}_1(t)$ and $\widetilde{\Omega}_2(t)$, the parameter $\lambda$, and the coupling strength $v$ between the cavity and the fiber on the final fidelity. Fig.~\ref{tu9} illustrates the dependence of the final fidelity on the relative errors of these parameters. As shown in Figs.~\ref{tu9}(a) and \ref{tu9}(b), the scheme exhibits robustness against fluctuations in the fitted pulses and in the parameters $\lambda$ and $v$.

In the experiment, Rydberg super atoms can be prepared by placing an ensemble of \(^{87}\rm{Rb}\) atoms in a magneto-optical trap in an ultra-high vacuum cavity. When the atomic ensemble is further cooled, the final atomic cloud can contain 25000 atoms \cite{b11}. Considering the experimental paramete \cite{zhang2010deterministic,t06,t07,t08}, Rydberg super-atoms containing \(n = 10^{4}\) \(^{87}\rm{Rb}\) atoms can be prepared in a magneto-optical trap. The control field drives the transition of the \(|e\rangle\leftrightarrow|r\rangle\) at a Rabi frequency \(\sim 2\pi\times10 \) MHz, he radiative decay rate of the intermediate state is \(\sim2\pi\times3\) MHz, the decay rate of the Rydberg state is \(\sim2\pi\times1\) kHz, and the decay of the cavity is \(\sim2\pi\times0.66\) MHz, Under this experimental condition, the fidelity of the scheme can reach more than $97.7\%$.

\section{Preparation of the W state of $N$ Rydberg superatoms}
\begin{figure}[h]
	\centering
	\includegraphics[width=1.0\linewidth]{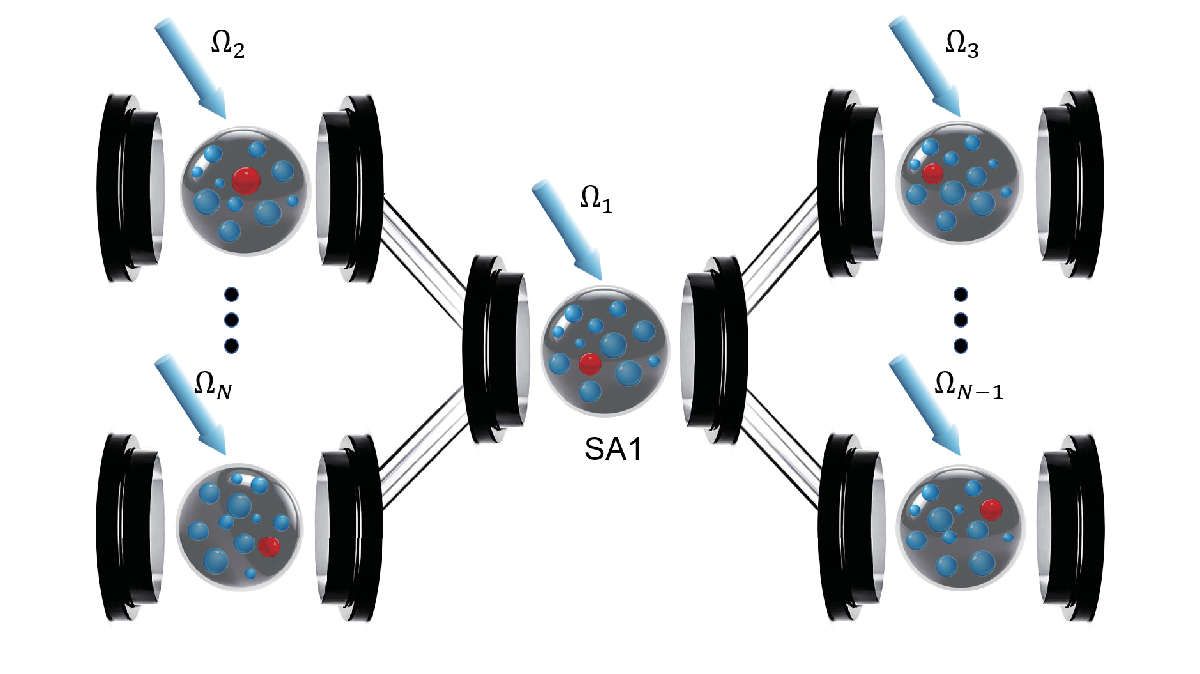}
	\caption{A schematic diagram illustrating the structure of $N$-Rydberg superatom-cavity system. Each of the Rydberg superatom is placed in a separate vacuum cavity,  with cavities 2 through $N$ all connected to cavity 1. $\Omega_N$ is the classical field-driven Rabi frequency in the $N$-th cavity.}
	\label{tu10}
\end{figure}

Because the more the number of entangled qubits is, the more significant the non-classical effect is, and the more important it is for quantum applications, the above W state preparation method will be further extended to the preparation of $N$ particle W state. The physical model is shown as Fig. \ref{tu10}, in which $N$ Rydberg superatoms are placed in different vacuum cavities, the second to $N$ cavities are connected to the first cavity, and the effective energy level structure of each superatom is the same as that of Fig. \ref{tu1} (b). The interaction Hamiltonian in the rotating wave approximation for this system is
\begin{align}
	H_{tot}&=H_{ac}+H_{al}+H_{cf},  \cr H_{ac}&=\sum^N_{k=1}\lambda_{k}a_{k}|E^{(1)}\rangle_k\langle{G}|+{\rm H.c.}, \cr   
	H_{al}&=\sum^N_{k=1}\Omega_{k}(t)|E^{(1)}\rangle_{k}\langle{R^{(1)}}|+{\rm H.c.}, \cr
	H_{cf}&=\sum^{N-1}_{k=1}v(a_{1}+a_{k+1})b_{k}^\dag+{\rm H.c.}.
\end{align}

Assuming that the initial state is $|\psi_1\rangle=|R^{(1)}GG\cdots GG\rangle$, and assuming that $v=\lambda$, $\Omega_k=\Omega (k=2, 3, 4, \cdots, N)$,  the final effective Hamiltonian can be obtained by using the same derivation method as Eq.~(\ref{Eq.1}) to Eq.~(\ref{Eq.8}).
\begin{align}
	H_{eff}=\Omega'_1(t) | \psi_1\rangle \langle \phi |+ \Omega'_2(t) | \phi\rangle  \langle \Psi | +\rm{H.c.},
	\label{Eq.21}
\end{align}
where$|\Psi\rangle=1/\sqrt{{N}-1}(|GR^{(1)}G\cdots GG\rangle+|GGR^{(1)}\cdots GG\rangle+\cdots +|GGG\cdots R^{(1)}G\rangle+|GGG\cdots GR^{(1)}\rangle)$, $\Omega'_1(t)=(N-1)/\sqrt{N^2-1}\Omega_1(t)$, 
$\Omega'_2(t)=1/\sqrt{N^2-1}\Omega(t)$. 
Comparing the Eq.(\ref{Eq.21}) and Eq. (\ref{Eq.5}) of the four-particle case, they are only different in the coefficients. It is obvious that the desired $N$ particle W state can be obtained by a process similar to the four-particle W state$1/\sqrt{N}|R^{(1)}GG\cdots GG\rangle+\sqrt{N-1}/\sqrt{N}|\Psi\rangle$. 
\section{Conclusion}

In this paper, we propose a scheme for rapid preparation of W state by applying the superadiabatic iteration technique to Rydberg superatomic system without precise control of experimental parameters and interaction time. A key feature of this scheme is that the anti-adiabatic Hamiltonian has the same form as the effective Hamiltonian, which simplifies the experimental implementation process and reduces the difficulty in experimental operation. The numerical simulation results show that the scheme can not only efficiently prepare the W state, but also has high fidelity and experimental feasibility. Numerical simulation analysis shows that the fidelity of the scheme can maintain high robustness even under the decoherence effect caused by atomic spontaneous emission and photon leakage. Even when the atomic spontaneous emission and photon leakage are large, the fidelity can still be maintained above 97.5\%, indicating the potential of the scheme in practice. In general, the preparation scheme of Rydberg super atom W state based on superadiabatic iteration technique proposed in this paper provides a new idea for the preparation of entangled States of multi-particle quantum systems in the future, and shows high scalability and practicability.

\bibliographystyle{apsrev4-2} 
\bibliography{apsrefdoi}

\end{document}